\documentclass[oupdraft]{bio}

\usepackage{amsfonts}
\usepackage{amsmath}
\usepackage{bm}
\usepackage{booktabs}
\usepackage{caption}
\usepackage{xcolor}
\usepackage{diagbox}
\usepackage{enumerate}
\usepackage{float}
\usepackage{graphicx}
\usepackage{lineno}
\usepackage{listings}
\usepackage{mathtools}
\usepackage{mdframed}
\usepackage{multicol}
\usepackage{multirow}
\usepackage{natbib}
\usepackage[normalem]{ulem}
\usepackage{pdfpages}
\usepackage{rotating}
\usepackage{setspace}
\usepackage{subcaption}
\usepackage{threeparttable}  
\usepackage{url}

\def\rmd{\mathrm{d}}
\def\trans{{}^\mathsf{T}}

\def\Pr{\mathrm{Pr}}
\def\trans{{}^\mathsf{T}}

\newcommand{\argmin}{\ensuremath{\operatornamewithlimits{arg\,min}}}

\newcommand{\iid}{\buildrel \mathrm{iid} \over\sim}
\newcommand{\ind}{\buildrel \mathrm{ind} \over\sim}
\newcommand{\tabincell}[2]{\begin{tabular}{@{}#1@{}}#2\end{tabular}}

\newtheorem{Pro}{Proposition}
\newtheorem{Exa}{Example}

\graphicspath{{figures/}}

\begin{document}

\title{Bayesian biclustering for microbial metagenomic sequencing data via multinomial matrix factorization}

\author{FANGTING ZHOU$^{1,2}$, KEJUN HE$^{2,\ast}$, QIWEI LI$^{3}$, ROBERT S. CHAPKIN$^{4}$, YANG NI$^{1,\ast}$ \\
$^{1}$Department of Statistics, Texas A\&M University, College Station, Texas, U.S.A. \\
$^{2}$Institute of Statistics and Big Data, Renmin University of China, Beijing, China \\
$^{3}$Department of Mathematical Sciences, The University of Texas at Dallas, Dallas, Texas, U.S.A \\
$^{4}$Department of Nutrition and Food Science, Texas A\&M University, College Station, Texas, U.S.A \\
{kejunhe@ruc.edu.cn, yni@stat.tamu.edu}}

\markboth
{Fangting Zhou and others}
{Bayesian biclustering via multinomial matrix factorization}	

\maketitle

\footnotetext{To whom correspondence should be addressed.}

\begin{abstract}
{High-throughput sequencing technology provides unprecedented opportunities to quantitatively explore human gut microbiome and its relation to diseases. Microbiome data are compositional, sparse, noisy, and heterogeneous, which pose serious challenges for statistical modeling. We propose an identifiable Bayesian multinomial matrix factorization model to infer overlapping clusters on both microbes and hosts. The proposed method represents the observed over-dispersed zero-inflated count matrix as Dirichlet-multinomial mixtures on which latent cluster structures are built hierarchically. Under the Bayesian framework, the number of clusters is automatically determined and available information from a taxonomic rank tree of microbes is naturally incorporated, which greatly improves the interpretability of our findings. We demonstrate the utility of the proposed approach by comparing to alternative methods in simulations.  An application to a human gut microbiome dataset involving patients with inflammatory bowel disease reveals interesting clusters, which contain bacteria families \emph{Bacteroidaceae}, \emph{Bifidobacteriaceae}, \emph{Enterobacteriaceae}, \emph{Fusobacteriaceae}, \emph{Lachnospiraceae}, \emph{Ruminococcaceae}, \emph{Pasteurellaceae}, and \emph{Porphyromonadaceae} that are known to be related to the  inflammatory bowel disease and its subtypes according to biological literature. Our findings can help generate potential hypotheses for future investigation of the heterogeneity of the human gut microbiome.}
{Bayesian nonparametric prior; Compositional data analysis; Mixture model; Phylogenetic Indian buffet process; Feature allocation.}
\end{abstract}

\maketitle

\section{Introduction} \label{introduction}

Microbes parasitic on various parts of the human body are inseparable from the well-being of their hosts. Recent studies have shown that microbiota have profound effects on the formation, development, and progression of numerous diseases like psoriasis \citep{benhadou2018psoriasis}, obesity \citep{castaner2018gut}, inflammatory bowel disease \citep[IBD,][]{franzosa2019gut}, preterm birth \citep{fettweis2019vaginal}, and diabetes \citep{tilg2014microbiota}. In this article, we focus on IBD, a chronic and complex disease that features heterogeneity at the microbiome level. As \cite{lloyd2019multi} pointed out, the disease activity is accompanied by molecular disruptions in microbial transcription, variations with taxonomic shifts, and other genomic activities. The seemingly strong association between gut microbes and IBD urges scientists to investigate microbial composition profiles in patients, which can improve our understanding of disease etiology and potentially lead to personalized treatments.

The emergence of high-throughput sequencing technology such as deep metagenomic sequencing has generated a plethora of data that have enabled researchers to quantitatively study both taxonomic and functional effect of microbiota on hosts \citep{turnbaugh2007human}. However, due to the compositional, sparse, heterogeneous, and noisy nature of the microbiome  abundance data, they pose serious challenges in statistical modeling.

{\it Composition.} Microbiome abundance data are inherently compositional \citep{gloor2017microbiome}, in the sense that individual counts are restricted by a sum constrain due to tissue size or sequencing depth. The abundance of each microbial component is only coherently interpretable relative to others within that sample. As a consequence, models that treat microbial taxa as independent variables may lead to substantial biases \citep{buccianti2013compositional}.

{\it Sparsity.} Microbial counts are sparse. Taking our IBD data as an example, more than 45\% of the observations are exact zeros, which greatly complicates the sampling distribution. Excessive zeros occur mainly for two reasons: (i) bacteria are not present in tested hosts and hence the zeros are true biological zeros, and (ii) the sequencing depth is not enough to capture rare bacteria which is referred to as technical zeros. Often, approaches need to explicitly differentiate between these two types of zeros to reduce estimation biases, which are addressed by the two-part model, the tobit model, and their combination \citep{liu2019statistical}. 

{\it Heterogeneity.} The composition of microbiota is heterogeneous and drastically different across hosts. Methods based on iid sampling are deemed unsuitable for microbiome data analysis. Individualized characterization is necessary to unravel genuine information and avoid spurious conclusions derived from a homogeneous modeling assumption.

{\it Noisiness.} Measurements from sequencing platforms contain high levels of noises due to the technical instability, which inevitably confounds with the biological variation that researchers strive to investigate. Methods that ignore the experimental noises are susceptible to false discoveries which will be propagated to downstream analysis and hinder scientific advancement.

Current statistical methodologies in the analysis of microbiome data are largely focused on the supervised learning framework. For example, in regression analysis where covariates are compositional, the linear log-contrast model with $\ell_1$ regularization was adopted in \cite{lin2014variable} and \cite{shi2016regression} to select relevant covariates in the analysis of metagenomic data. However, they do not explicitly take into account the excessive zeros but replace them with arbitrary small numbers. When treating compositional data as response, a sparse Dirichlet-multinomial regression model was employed in \cite{chen2013variable} to associate microbiome composition with environmental covariates. The method is able to account for over-dispersion of observed counts and select important covariates. \cite{xia2013logistic} introduced an additive logistic normal multinomial regression model and selected significant covariates via a group $\ell_1$ penalty. \cite{chen2016two} proposed a zero-inflated Beta regression model. The model includes a logistic regression component to model presence or absence of microbes in samples and a Beta regression component to model non-zero microbiome abundance. \cite{wadsworth2017integrative} developed a Bayesian Dirichlet-multinomial regression model combined with spike-and-slab priors to select important covariates that are predictive of microbial abundances. \cite{grantham2020mimix} proposed a Bayesian mixed-effects model for	capturing the effects of treatment,  covariates, and latent factors on microbial responses.

There are also a rising number of models focusing on revealing microbiome interactions. For example, \cite{friedman2012inferring} proposed to estimate the Pearson correlations between log-transformed components of compositional data under the assumption of sparsity, which is later implemented more efficiently with parallel computing by \cite{watts2018fastspar}. A composition-adjusted thresholding was proposed by \cite{cao2019large} to obtain a sparse correlation estimate. More recently, \cite{cai2019differential} developed a Markov random field model to detect differential microbial networks. A key step in their approach is to dichotomize microbial compositions into a binary matrix. However, the dichotomization in their approach is based on a fixed quantile, the choice of which is somewhat arbitrary and sensitive.

While the majority of microbiome data analyses are performed in a supervised manner, in this paper we focus on an unsupervised learning task, namely, the probabilistic matrix factorization of microbiome data which can also be interpreted as overlapping biclustering. Many matrix factorization techniques have been proposed to handle continuous matrices \citep{bhattacharya2011sparse, rovckova2016fast}, non-negative matrices \citep{lee2000algorithms, hoyer2004non}, count matrices \citep{zhou2012beta, gopalan2014bayesian}, and binary matrices \citep{meeds2007modeling, ni2019bayesian, wu2019barlcm}. However, none of these methods is directly applicable to compositional microbiome data. To account for both sparsity and heterogeneity of microbiome data, a matrix factorization approach based on Dirichlet prior and low dimensional representation was proposed in \cite{shafiei2015biomico}. \cite{ren2017bayesian} developed a Bayesian nonparametric ordination approach to capture the high-dimensional microbial dependencies via low-dimensional latent factors. Recently, a low rank approximation method was proposed by \cite{cao2019multisample} which minimizes the multinomial likelihood-based loss function combined with a nuclear norm regularization on the composition matrix. They focused on recovering the composition and matrix factorization rather than inferring latent clustering structure which is the main objective of this paper. \cite{xu2020zero} developed a zero-inflated Poisson factor model with Poisson rates negatively related to inflated zero occurrences.  Again, their main focus was on reducing the dimensionality of the microbiome data and a separate clustering algorithm is required to identify the clusters. 

In this paper, we propose a Bayesian multinomial matrix factorization (MMF) model that infers the latent clustering structure from compositional, sparse, heterogeneous, and noisy microbiome data. The proposed MMF introduces a mixture model representation of observations through a set of latent variables to indicate the relative abundance of taxa. In essence, this simple formulation of the sampling model \textit{adaptively} dichotomizes the multinomial observations into a binary matrix, which is more robust to noise and does not require a separate treatment of excessive zeros. Given the binary indicator matrix, priors are imposed hierarchically to characterize the heterogeneity via latent features. Specifically, we construct the hierarchical model with a combination of latent logit model, phylogenetic Indian buffet process prior \citep[pIBP,][]{miller2008the,chen2016posterior}, and beta-Bernoulli prior. Using pIBP, we are able to infer an unknown number of overlapping clusters/communities of the taxa. pIBP also takes into account the taxonomic relationships among the taxa, which gives rise to more interpretable and reliable results. Conditional on the clusters of taxa, the beta-Bernoulli prior are assigned to cluster hosts, again allowing overlaps. Moreover, the sparse nature of the pIBP and beta-Bernoulli priors leads to an identifiable matrix factorization under a mild condition. Using simulations, we demonstrate that the proposed MMF has favorable performance compared to competing methods and is relatively robust to the choice of hyperparameters and misspecified tree information. We then apply MMF to an IBD microbiome dataset \citep{qin2010human}, which  reveals interesting clusters containing bacteria families \emph{Bacteroidaceae}, \emph{Bifidobacteriaceae}, \emph{Enterobacteriaceae}, \emph{Fusobacteriaceae}, \emph{Lachnospiraceae}, \emph{Ruminococcaceae}, \emph{Pasteurellaceae}, and \emph{Porphyromonadaceae} that are known to be related to the IBD and its subtypes according to biological literature. Despite the exploratory nature of this study, our findings can help generate hypotheses for further investigation of the heterogeneity of the human gut microbiome.

The rest of this paper is organized as follows. We introduce the proposed MMF model in Section \ref{model}. Posterior inference based on Markov chain Monte Carlo (MCMC) sampling is described in Section \ref{posterior}. In Sections \ref{simulation} and \ref{data}, we respectively illustrate our approach with simulation studies and the analysis of an IBD dataset. This paper is concluded with a brief discussion in Section \ref{discussion}.

\section{Model} \label{model}

\subsection{Classifying Taxon Abundance via Adaptive Dichotomization } \label{sampling}

Let $x_{ij}$ denote the observed count of taxon $j$ in host $i$, $j = 1, \dots, p$ and $i = 1, \dots, n$. Let $\bm{x}_i = (x_{i1}, \ldots, x_{ip})\trans$ and $N_i = \sum_{j =1 }^p x_{ij}$. We assume $\bm{x}_i$ follows a Dirichlet-multinomial distribution, 
\begin{equation*}
\bm{x}_i \sim \mathrm{Multinomial}(N_i, \bm{\pi}_i)
\end{equation*}
with host-specific relative abundances, $$\bm{\pi}_i = (\pi_{i1}, \ldots, \pi_{ip})\trans \sim \mathrm{Dirichlet}(\bm{\eta}_i),$$ where $\bm{\eta}_i = (\eta_{i1}, \dots, \eta_{ip})\trans$. Note that the Dirichlet-distributed relative abundances $\bm{\pi}_i$ can be equivalently represented as normalized gamma random variables $\bm{\pi}_i = \bm{\gamma}_i /  \sum_{j = 1}^p \gamma_{ij}$ with unnormalized relative abundances $\bm{\gamma}_i = (\gamma_{i1}, \dots, \gamma_{ip})\trans$ and $\gamma_{ij} \ind \mbox{Gamma}(\eta_{ij}, 1)$, where the gamma distribution is parameterized as $\mbox{Gamma}(x; a, b) = \frac{b^a}{\Gamma(a)} x^{a - 1} e^{- bx}.$

We introduce a latent indicator variable $z_{ij}$ to classify whether a taxon $j$ is significantly \textit{present} or \textit{absent} in host $i$. However, there is no consensus on the classification of taxa based on absolute or relative abundances. In supervised tasks, the classification rule may be chosen to minimize certain objective functions. For example, when the goal is to predict a response variable with microbiome covariates, one can potentially find an optimal dichotomization that minimizes the prediction error. Lack of such gold standard in matrix factorization, we propose a mixture model to probabilistically classify raw taxa counts into states of high presence versus low presence and absence,
\begin{equation} \label{mgam}
\gamma_{ij} \sim I(z_{ij} = 1) \mbox{Gamma}(s_j, 1) + I(z_{ij} = 0) \mbox{Gamma}(t_j, 1) \mbox{~~with~~} s_j > t_j.
\end{equation}
In words, due to the constraint $s_j > t_j$, $z_{ij} = 1$  indicates high relative abundance and $z_{ij} = 0$ indicates low relative abundance. The choice of the two-component mixture model is motivated by the fact that the distribution of microbial abundances tend to be overdispersed and bi-modal \citep{koren2013guide,lahti2014tipping}. 
If we further constrain $t_j<1$, the prior \eqref{mgam} becomes a spike-and-slab prior with Gamma$(t_j,1)$ as the spike distribution, assigning an infinite mass at zero. Through the multinomial sampling and adaptive discretization, the zero counts would naturally fall into the category of low abundance with high probability. Therefore, we do not need an extra zero-inflated component to explicitly deal with the zero counts in our model. In addition, the adaptive dichotomization also accounts for sequencing errors as in \cite{parmigiani2002statistical}.
The induced distribution of $\bm{x}_i$ is a discrete mixture of Dirichlet-multinomial distributions with $2^p$ components, with each component corresponding to one configuration of $(z_{i1}, \dots, z_{ip})\trans$. The latent variable $z_{ij}$ can be viewed as a denoised version of the raw observations $x_{ij}$. A similar idea of denoising was recently used by \cite{cai2019differential} where they assumed that taxa with relative abundances lower than 0.001\% are due to noise or sequencing errors, and adopted the 0.25 quantile as a hard cutoff for more abundant taxa. Our approach differs from theirs in that we do not need to fix a cutoff and the proposed method adaptively dichotomizes the data. We assign hyperpriors on the unknown parameters $(s_j, t_j)$,
\begin{equation} \label{hyper}
p(s_j, t_j) = \mathrm{Gamma}(s_j | \alpha_s, \beta_s) \times \mathrm{Gamma}(t_j | \alpha_t, \beta_t) \times I(s_j>t_j),
\end{equation}
with $\alpha_s = \alpha_t = \alpha = 1$ and $\beta_s = \beta_t = \beta = 0.1$. Sensitivity analyses will be performed on the choice of all the hyperparameters in Section \ref{simulation} and Supplementary Material (Section B).

The mixture model in \eqref{mgam} can reliably classify taxa with well separated relative abundances into two states. However, the classification can have greater uncertainties for taxa with less variable relative abundances across observations. In Section \ref{binary}, we will introduce latent structures on $\bm{Z} = (z_{ij})$  that stabilize uncertain classifications, reduce the dimensionality, and induce overlapping cluster structure for both hosts and microbial taxa.

\subsection{ Biclustering Taxa and Hosts via Binary Matrix Factorization \label{binary}}

We introduce lower-dimensional matrices to characterize the heterogeneity of both rows and columns of $\bm{Z}$. In particular, we let $\bm{A} = (a_{ik}) \in \{0, 1\}^{n \times K}$ and $\bm{B} = (b_{jk})  \in \{0, 1\}^{p \times K}$ denote the host-cluster and taxon-cluster matrices with $K$ clusters. The clustering interpretations of $\bm{A}$ and $\bm{B}$ will be elaborated in Section \ref{prior}. The number $K$ of columns of $\bm{A}$ and $\bm{B}$ is usually much smaller than the dimensions of the original data ($n$ and $p$). We link $\bm{A}$ and $\bm{B}$ to $z_{ij}$  by a latent logit model 
\begin{equation} \label{logistic}
\mathrm{logit}\{\Pr(z_{ij} = 1)\} = c_j + \sum_{k = 1}^K a_{ik} w_{jk} b_{jk},
\end{equation}
where $\mathrm{logit}(p) = \log\{p / (1 - p)\}$. If a group of hosts have a common activated biological pathway (related to normal body functions or diseases) that involves a common set of taxa, then these taxa are likely to have significant presence in those host samples. Therefore, we choose to constrain $w_{jk}$ to be positive, although in principle they can take any values; similar considerations in a different context were made in \cite{wood2006a}. Parameter $c_j$ represents the log odds ratio of baseline probability of the presence of taxon $j$. We assume weakly informative priors on $w_{jk}$ and $c_j$,
$w_{jk} \sim \mathrm{Gamma} (\alpha_w, \beta_w)$, and $c_j \sim \mathrm{N}(\mu_c, \sigma_c^2)$, with $\mu_c = 0$, $\sigma_c^2 = 100$, $\alpha_w  = 1$, and $\beta_w = 0.1$.

\subsection{Indian Buffet Process and Taxonomic Rank Tree} \label{prior}

The host-cluster matrix $\bm{A}$ and taxon-cluster matrix $\bm{B}$ can be interpreted as clustering of rows and columns of $\bm{Z}$, respectively. Host $i$ (taxon $j$) belongs to cluster $k$ if the corresponding $a_{ik} = 1$ ($b_{jk} = 1$). Since we do not constrain $\bm{A}$ and $\bm{B}$ to having unit row sums, clusters can have overlaps. This is useful in microbiome applications because a taxon can be active in multiple communities and likewise a host can also belong to more than one group. To make inference on these two matrices, we will impose a Bayesian nonparametric prior on $\bm{B}$ that can automatically determine the number $K$ of clusters.

The Indian buffet process \citep[IBP,][]{griffiths2005infinite} has been widely used as a Bayesian nonparametric prior on binary matrices with potentially unbounded number of columns. IBP assumes the rows of the binary matrix are exchangeable. This assumption becomes a limitation when the rows (taxa) are seemingly dependent as in our case. For instance, the relationships between taxa are commonly organized as a taxonomic rank tree. Taxa with smaller distances on the tree tend to have similar biological functions and therefore are expected to have higher probability of being in the same cluster. To incorporate this prior knowledge, we adopt the phylogenetic IBP \citep[pIBP,][]{miller2008the} to encourage taxonomically similar taxa to form clusters.

To describe the generating process of pIBP, we first assume a fixed and finite number $\widetilde{K}$ of clusters and will later relax it. Conditional on $\widetilde{K}$, we associate a parameter $p_k$ to each column of $\bm{B}$, which is assigned a $\mathrm{Beta}(m / \widetilde{K}, 1)$ prior. We put a $\mathrm{Gamma}(1, 1)$ prior on $m$ to infer its value from data. While the columns of $\bm{B}$ are still independent as in IBP, entries within each column are generated jointly, with the pattern of dependence characterized by a stochastic process on a taxonomic rank tree. The tree has $p$ taxa of interest as leaves and higher taxonomic ranks as internal/root nodes. Assume the path from every leaf up to the root contains $(L - 1)$ internal nodes, and each edge has length $1 / L$ so that the total length of the path from every leaf  to the root is 1. This implies that the marginal prior probability of $b_{jk} = 1$ is the same across taxa $j = 1, \dots, p$. 

To generate the entries of the $k$th column, we proceed as follows: i) assign value zero to the root node of the tree; ii) along any path from the root to a leaf, let the value change to one with an exponential rate $- \log(1 - p_k) / L$; iii) once the value has changed to one along a path from the root, all leaves below that change point are assigned value one; and iv) set the entries in the $k$th column of $\bm{B}$ to the values of the corresponding leaves. By construction, leaves that are closer on the tree tend to receive identical values (of zeros or ones) in each column and therefore the corresponding taxa are more likely to fall in the same cluster. Note that the marginal prior probability of $b_{jk} = 1$ is $p_k$, as given in the original paper of \cite{miller2008the}. To remove the dependency of the generating process from a fixed $\widetilde{K}$, we let $\widetilde{K}$ go to infinity and obtain the pIBP. Hereafter we omit empty columns and denote the number of non-empty columns by $K$.

\begin{figure}
\centering
\includegraphics[width = 0.8 \textwidth]{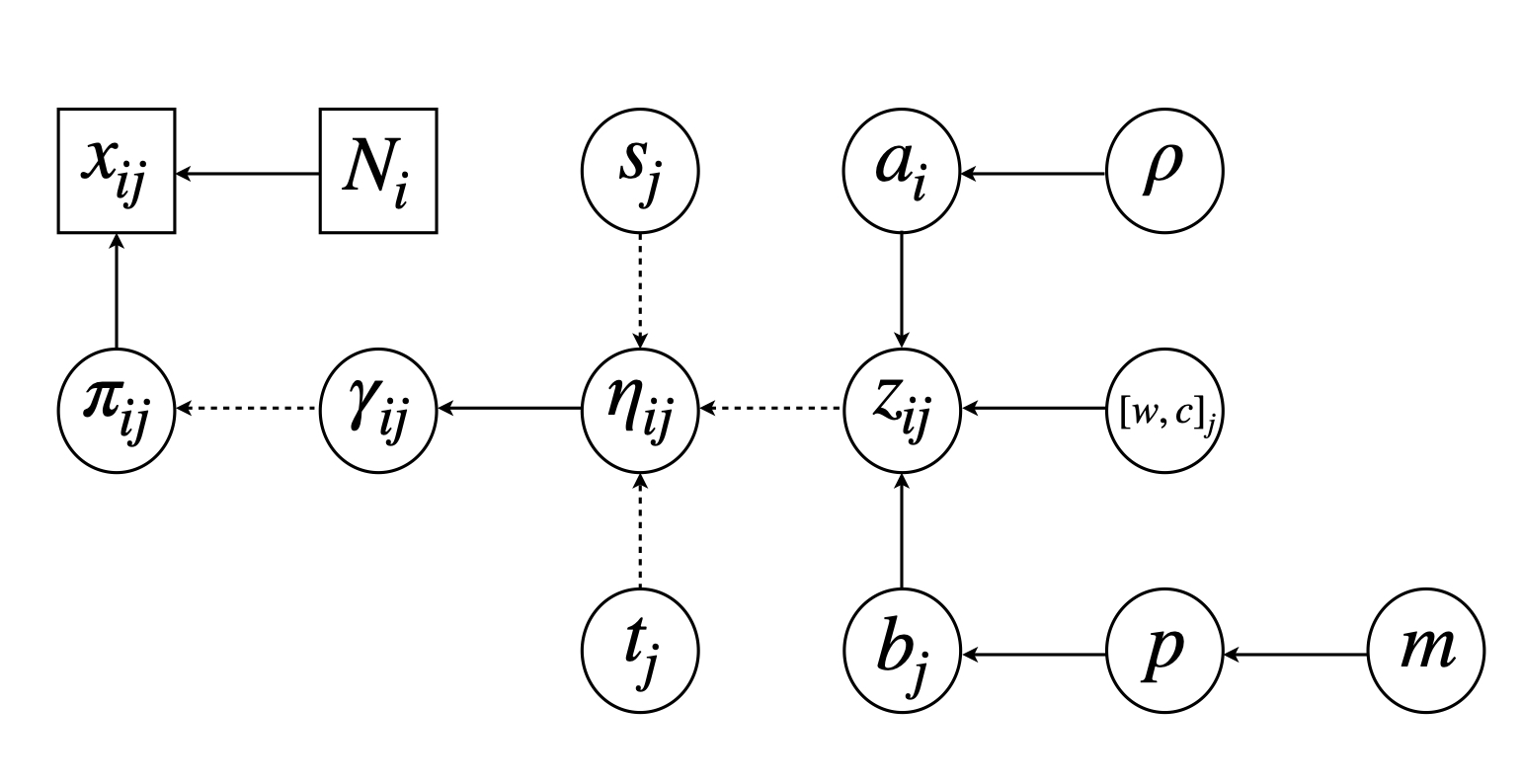}
\caption{A graphical representation of the model. Dashed edges and squares are deterministic, and solid edges and circles are stochastic.}
\label{graph}
\end{figure}

Conditional on taxon-cluster matrix $\bm{B}$ (only via $K$), each element $a_{ik}$ in $\bm{A}$ follows an independent beta-Bernoulli distribution $a_{ik} \iid \mathrm{Bernoulli}(\rho)$, and $\rho \sim \mathrm{Beta}(\alpha_\rho, \beta_\rho)$ with $\alpha_\rho = \beta_\rho = 1$. The complete hierarchical model is represented as directed acyclic graph in Figure \ref{graph}.

Our choice of the nonparametric pIBP prior allows for flexible modeling of the latent structures. First, the number of clusters is potentially unbounded (i.e., it can increase as the sample size grows) and can be inferred from data. Second, given the number of clusters, the prior model assigns positive mass on any taxon-cluster matrix $\bm{B}$ (via pIBP) and any host-cluster matrix $\bm{A}$ (via independent Bernoulli's conditional on $\bm{B}$). Third, through the logit link, the prior model on $\bm{A}$ and $\bm{B}$ also induces a flexible prior on the latent abundance matrix $\bm{Z}$.

\subsection{Identifiability of the Proposed Model}

Matrix factorization is often non-identifiable without additional assumptions.  For example, model \eqref{logistic} can be written in a matrix form,
\begin{align*}
\bm{Q}=\bm{C}+\bm{A}\bm{B}\trans,
\end{align*} 
where $\bm{Q}=(q_{ij})$ with $q_{ij}=\mathrm{logit}\{\Pr(z_{ij} = 1)\}$, $\bm{C}=\bm{1}_p\bm{c}\trans$ with $\bm{1}_p=(1,\dots,1)\trans$ and $\bm{c}=(c_1,\dots,c_p)\trans$, and, slightly abusing the notation, $\bm{B}=(w_{jk}b_{jk})$ absorbs the weights $w_{jk}$. Let $\widetilde{\bm{A}}=\bm{A}\bm{P}$ and $\widetilde{\bm{B}}=\bm{B}\bm{P}$ for any $K\times K$ orthogonal matrix $\bm{P}$.   It is obvious that $\widetilde{\bm{A}}\widetilde{\bm{B}}\trans=\bm{A}\bm{P}\bm{P}\trans\bm{B}\trans=\bm{A}\bm{B}\trans$. Consequently, $(\bm{A},\bm{B})$ and $(\widetilde{\bm{A}},\widetilde{\bm{B}})$ would lead to the same $\bm{Q}$ and the same sampling distribution, and are therefore non-identifiable in general. However, the fact that $\bm{A}$ is binary makes the proposed matrix factorization identifiable up to column permutations under a mild condition.

\begin{Pro} If $\bm{A}$ is a binary matrix and there exists an integer matrix $\bm{R} \in \mathbb{Z}^{K \times n}$ such that $\bm{R} \bm{A} = \bm{I}$, then $\bm{A}$ and $\bm{B}$ is uniquely identifiable up to column permutation.
 
\noindent \emph{Proof.} Let $\widetilde{\bm{A}} = \bm{A} \bm{P}$ with an orthogonal matrix $\bm{P}$.
We will show that  $\bm{P}$ must be a permutation matrix if $\widetilde{\bm{A}}$ is a binary matrix. 
We have 
$$\bm{R} \widetilde{\bm{A}}=\bm{R} \bm{A} \bm{P} = \bm{P}.$$	
Since both $\bm{R}$ and $\widetilde{\bm{A}}$ are integer matrices, $\bm{P}$ must be an integer matrix. This implies that each row of $\bm{P}$ is a unit vector and $\bm{P}$ is therefore a permutation matrix. $\square$
\end{Pro}

The condition is, in our opinion, mild. For example, it is satisfied if for any  $k=1,\dots,K$, there exists $i=1,\dots,n$ such that $\bm{a}_i = \bm{e}_k$ where $\bm{a}_i$ is the $i$th row of $\bm{A}$  and $\bm{e}_k$ is a unit vector with 1 at its $k$th entry (in this case $\bm{R}$ would simply be a binary matrix that acts to select those $K$ rows of $\bm{A}$). In words, the proposed model is identifiable if for any cluster $k$, there exists at least one member of this cluster that does not belong to any other clusters.
Below, for completeness, we give a non-identifiable example when the condition of Proposition 1 is not met.

\begin{Exa}
Suppose
\begin{align*}
\bm{A} \bm{P} = \begin{pmatrix}
0 & 0 & 1 & 1 \\
1 & 1 & 0 & 0 \\
0 & 1 & 0 & 1 \\
1 & 0 & 0 & 1
\end{pmatrix} \begin{pmatrix}
1/2 & 1/2 & 1/2 & -1/2 \\
1/2 & 1/2 & -1/2 & 1/2 \\
1/2 & -1/2 & 1/2 & 1/2 \\
-1/2 & 1/2 & 1/2 & 1/2
\end{pmatrix} = \begin{pmatrix}
0 & 0 & 1 & 1 \\
1 & 1 & 0 & 0 \\
0 & 1 & 0 & 1 \\
0 & 1 & 1 & 0
\end{pmatrix} = \widetilde{\bm{A}}.
\end{align*}
 Then the matrix $\bm{P}$ satisfies $\bm{P} \bm{P}\trans = \bm{I}$ but is not a permutation matrix.
\end{Exa}

\section{Posterior Inference} \label{posterior}

The proposed MMF is parameterized by $\left\{\bm{A},\bm{B}, \bm{Z}, \{\bm{\gamma}_i\}_{i = 1}^n \{\bm{w}_j, c_j, s_j, t_j\}_{j= 1}^p, \{p_k\}_{k = 1}^K, m, \rho\right\}$. We carry out the posterior inference by MCMC simulation. To improve mixing, we marginalize out unnormalized relative abundance parameters $\bm{\gamma}_i$'s. While other parameters are trivial to update with Gibbs or Metropolis-Hasting (M-H), care must be taken in updating $\bm{B}$ and $\{p_k\}_{k = 1}^K$, details of which are provided below. The updating procedures of other parameters are presented in the Supplementary Material (Section A). We let $\bm{b}_k$ and $\bm{b}_k^{-j}$ respectively denote the $k$th column of $\bm{B}$ and the $k$th column of $\bm{B}$ without the $j$th entry. Sequentially for $j = 1, \dots, p$, we cycle through the following three steps.
\begin{enumerate}[Step i.]
\item Update existing (non-empty) columns $k = 1, \ldots, K$ of $\bm{B}$. For $j = 1, \ldots, p$, we sample the binary $b_{jk}$ from the full conditional distribution,
\begin{align*}
p(b_{jk} | \cdot) \propto
p(b_{jk} | \bm{b}_k^{-j}, p_k) \prod_{i = 1}^n p(z_{ij} | \{a_{ik}, b_{jk}, w_{jk}\}_{k = 1}^K, c_j).
\end{align*}
While an analytic form of $p(b_{jk} | \bm{b}_{k}^{-j}, p_k)$ is hard to obtain, it can be computed by the sum-product algorithm exactly and efficiently. Details of the sum-product algorithm can be found in \cite{bishop2006pattern}. Importantly, if  a column becomes all zeros after update, we delete that column and reduce $K$ by 1.
\item Update $p_k$ for existing columns of $\bm{B}$. Suppose $b_{jk}$ is any non-zero entry in the $k$th column. The full conditional of $p_k$ is given by
\begin{align*}
p(p_k | \bm{b}_k, m) \propto p(p_k | b_{jk}, m) \, p(\bm{b}_k^{-j} | p_k, b_{jk}), 
\end{align*}
where the first factor is a standard uniform distribution \citep{miller2008the} and the second factor can be efficiently computed by decomposing it into a series of univariate conditional distributions using the chain rule. For example, without loss of generality, assuming $j=1$, then 
$$p(\bm{b}_k^{-1} | p_k, b_{1k})=p(b_{2k}|p_k,b_{1k}) \,  p(b_{3k}|p_k,b_{1k},b_{2k})\cdots \,  p(b_{pk}|p_k,b_{1k},\dots,b_{p-1,k}),$$ 
 where each factor can be computed again using the sum-product algorithm. Since we only know the full conditional up to a normalization constant, we draw $p_k$ by a M-H step, where a new value is proposed from $p_k^*\sim q(p_k^*|p_k) = \mathrm{N}(p_k, \sigma_k^2)$ and is accepted with probability
\begin{align*}
\min \left\{1, \frac{q(p_k | p_k^*) \,  p(\bm{b}_k^{-j} | p_k^*, b_{jk})}{q(p_k^* | p_k) \,  p(\bm{b}_k^{-j} | p_k, b_{jk})} \,  I(p_k^* \in [0, 1])\right\}.
\end{align*}
Following the default choice in \cite{miller2008the}, we choose $\sigma_k^2 = c p_k (1 - p_k) + \delta$,  with $c = 0.06$ and $\delta = 0.08$.
\item Propose new columns. After all the existing columns are updated, we propose to add new columns. We first draw
$$
K^* \sim \mathrm{Poisson}\left(m \left\{\psi \left((P - 1) / L + 1\right) - \psi \left((P - 2) / L + 1\right) \right\}\right),
$$
where $\psi(\cdot)$ is the digamma function and $P$ is the total number of nodes in the tree. If $K^* = 0$, we will go to the next step. Otherwise, we propose a set of new parameters $\bm{a}_k^* = (a_{1k}^*, \ldots, a_{nk}^*)\trans$ and $w_{jk}^*$ from their prior distributions, $k = K + 1, \ldots, K + K^*$. We accept new columns and the associated new parameters with probability
\begin{align*}
\min \left\{1, \frac{\prod_{i = 1}^n p \big(z_{ij} | \{a_{ik}, b_{jk}, w_{jk}\}_{k = 1}^{K}, \{a_{ik}^*, b_{jk}^*, w_{jk}^*\}_{k = K + 1}^{K + K^*}, c_j \big) }{\prod_{i = 1}^n p \big( z_{ij} | \{a_{ik}, b_{jk}, w_{jk}\}_{k = 1}^{K}, c_j \big)} \right\},
\end{align*}
where $b_{j, K + 1} = \ldots = b_{j, K + K^*} = 1$. Lastly, if new columns are accepted, we increase $K$ by $K^*$ and  sample $p_k$ for the new columns by a M-H step,
\begin{equation*}
p(p_k | \bm{b}_k) \propto \{1 - (1 - p_k)^{1 / L}\} (1 - p_k)^{(P - 2) / L} / p_k.
\end{equation*}
\end{enumerate}

To summarize the posterior distribution based on the Monte Carlo samples, we proceed by first calculating the maximum a posteriori estimate $\widehat{K}$ of $K$ from the marginal posterior distribution. Conditional on $\widehat{K}$, we find an estimate of $\bm{B}$ by the following procedure. For any matrices $\bm{B}$, $\widetilde{\bm{B}} \in \{0, 1\}^{p \times \widehat{K}}$, we define a distance
\begin{equation} \label{eqn:distB}
d(\bm{B}, \widetilde{\bm{B}}) = \min_\pi H(\bm{B}, \pi(\widetilde{\bm{B}})),
\end{equation}
where $\pi(\widetilde{\bm{B}})$ denotes a permutation of the columns of $\widetilde{\bm{B}}$ and $H(\cdot, \cdot)$ is the Hamming distance between two binary matrices, i.e., counting the number of different entries between the two matrices. A point estimator $\widehat{\bm{B}}$ of $\bm{B}$ is then obtained as
\begin{equation*}
\widehat{\bm{B}}= \argmin_{\widetilde{\bm{B}}} \int d(\bm{B}, \widetilde{\bm{B}}) \, \rmd p(\bm{B} | \cdot),
\end{equation*}
where $p(\bm{B} | \cdot)$ denotes the marginal posterior distribution of $\bm{B}$ given $\widehat{K}$. Empirically, both the integration and the optimization can be approximated using the available Monte Carlo samples. Specifically, we define the posterior mode $\widehat{\bm{B}}$ as $$\widehat{\bm{B}} = \argmin_{\widetilde{\bm{B}} \in \mathcal{B}} \frac{1}{S} \sum_{s = 1}^S d(\bm{B}^{(s)}, \widetilde{\bm{B}}),$$ where $\mathcal{B} = \{\bm{B}^{(s)}, s = 1, \ldots, S\}$ is the set of posterior samples of $\bm{B}$ and the distance function is given in \eqref{eqn:distB}. Conditional on $\widehat{\bm{B}}$, we continue to run the Markov chain for a while. Then the point estimates of other parameters are obtained as the posterior means computed from the new Monte Carlo samples.

\section{Simulation} \label{simulation}

In our simulation study, we considered a dataset with $n = 300$ hosts, $p = 46$ taxa, and $K = 6$ true clusters; similar in size to the later application. For $k = 1, \ldots, K$, we first set $a_{ik} = 1$, $i = 50 (k - 1) + 1, \ldots, 50 k$, and $0$ all the others. Then we randomly changed 10\% of zero entries in the host-cluster matrix $\bm{A}$ to one. We used the same taxonomic rank tree as in later application to generate the taxon-cluster matrix $\bm{B}$, which had $L + 1= 5$ levels. Furthermore, cluster-specific probability parameters $p_k$ were all set to 0.3. The resulting true $\bm{A}$ and $\bm{B}$, along with the phylogenetic tree are shown in Figure S.1 in the Supplementary Material. By construction, each taxon or host was allowed to belong to multiple clusters. Latent indicators $z_{ij}$ were generated from the logit model \eqref{logistic} with $\bm{w}_j = \bm{w} = (2.0, 2.5, 3.0, 3.5, 4.0, 4.5)\trans$ and $c_j = \log 0.5$. For the unnormalized relative abundance $\gamma_{ij}$, we simulated them from the gamma mixture model \eqref{mgam} with varying degrees of separation of mixture components, $(s_j, t_j) = (s, t) = (2, 0.7)$, $(3, 0.6)$, and $(5, 0.5)$. Among these three simulation scenarios, $(s, t) = (2, 0.7)$ was the most difficult as it induced the least separation between the two mixture component. The observations were finally generated from the multinomial sampling model for which the total counts were drawn from the discrete uniform distribution $\mathrm{U}(50, 500)$. 

We ran the MCMC algorithm of MMF for 5,000 iterations with 10 random initial clusters. The first 2,500 iterations were discarded as burn-in and posterior samples were retained every 5th iteration after burn-in. On average, it took 3.8 hours  on a 2.3 GHz Quad-Core Intel Core i7 laptop. To evaluate the recovery accuracy, we calculated the estimation errors for both $\bm{A}$ and $\bm{B}$. Specifically, we computed the Hamming distance between the estimated and true $\bm{A}$ and $\bm{B}$, normalized by the respective total number of elements. When the estimated number of clusters was different from the truth, we padded the smaller matrix with columns of zeros, making the resulting matrices comparable in dimension.

\noindent \textbf{Method evaluation.} The results under three sets of true values of $(s, t)$ are summarized in Table \ref{t1} based on 50 repeated simulations. As expected, the performance improved as the two mixture components in \eqref{hyper} became more separated, from $(s, t)=(2, 0.7)$ to $(5, 0.5)$. The proposed MMF was able to identify the correct number $K$ of clusters at least 95\% of the time. Figure S.1 in the Supplementary Material depicts the estimated host-cluster and taxon-cluster matrices $\widehat{\bm A}$ and $\widehat{\bm B}$ of the proposed MMF from one simulation result with the worst error rate in the scenario $(s, t) = (5, 0.5)$ after adjusting for label switching and dropping redundant columns. They are visually quite close to the truth, indicating that the proposed method was able to consistently and accurately identify the clusters of hosts and taxa.

\noindent \textbf{Comparisons with competing methods.} Matrix factorization has been studied extensively in the literature. We compared the proposed MMF with three existing alternative matrix factorization methods, the low rank approximation (LRA, \citealt{cao2019multisample}), the non-negative matrix factorization (NNMF, \citealt{cai2017learning}), and the zero-inflated Poisson factor model (ZIPFM, \citealt{xu2020zero}). 

In order to compare the performance of biclustering, the overlapping clustering method, fuzzy c-means \citep{bezdek1984fcm}, was applied to the latent factors or low rank matrices obtained from the competing methods. The dimension of latent factors was chosen by their default optimization procedure. The number of clusters was set to the truth $K=6$ for competing methods whereas it was estimated for the proposed MMF.  The estimation errors were defined by first converting clustering results to binary host-cluster or taxon-cluster matrix, and then calculating the distance between the estimated and true matrices as was done for the proposed method. In addition, we also considered a two-step approach that was similar to the proposed MMF, denoted by TSMF. Specifically, instead of joint modeling, the two-step approach first dichotomized observations using a default cutoff as suggested in \cite{cai2019differential} and then applied the same Bayesian nonparametric binary matrix factorization method as in MMF to the binary data. The results of all the methods are reported in Table \ref{t1}. The proposed MMF consistently outperformed the competing methods in most settings (keeping in mind that the number of clusters was set to truth for LRA, NNMF, and ZIPFM), especially for the taxon-cluster matrix $\bm{B}$. Although when $(s,t)$ was specified to be $(2, 0.7)$, the estimation error of $\bm{A}$ in LRA was smaller, the significantly more accurate result of estimating $\bm{B}$ in the proposed MMF shows the benefit of using the phylogenetic tree information.

\noindent \textbf{Additional simulations with different values of $\bm{w}$.} We performed additional simulation for $\bm{w} = (0.5, 0.6, 0.7, 0.9, 1.1, 1.2)\trans$ and $(1.0, 1.2, 1.5, 1.7, 2.0, 2.3)\trans$ in the Supplementary Material (Section B), which led to similar conclusion as above.

\noindent \textbf{Misspecified model.} For fairer comparison, we mimicked the generating process of microbial metagenomic sequencing data, which was different from the proposed model. In this experiment,  $\bm{A}$ and $\bm{B}$ were first generated as before. Then, the true counts $\bm{Y} = (y_{ij})$ were simulated from a negative binomial model,
\begin{align*}
\mathrm{NB}(y_{ij}; \mu_{ij}, \kappa_{ij}) = \frac{\Gamma(\kappa_{ij} + y_{ij})}{\Gamma(\kappa_{ij}) y_{ij}!} \left(\frac{\kappa_{ij}}{\kappa_{ij} + \mu_{ij}}\right)^{\kappa_{ij}} \left(\frac{\mu_{ij}}{\kappa_{ij} + \mu_{ij}}\right)^{y_{ij}},
\end{align*}
where $\mu_{ij} = \kappa_{ij} = \exp(\sum_k a_{ik} w_{jk} b_{jk} + c_j)$ such that $\mathbb{E}(y_{ij})=\mu_{ij}$ and $\mathrm{Var}(y_{ij})=2\mu_{ij}$. The counts were  then proportionally down-sampled from a multinomial distribution with sequencing depth uniformly chosen from $(50, 500)$. We subsequently applied the proposed MMF and the competing methods to the down-sampled dataset. The results based on 50 repetitions are summarized in Table \ref{t2}, which show the overall competitive performance of the proposed MMF over the alternatives, especially in estimating $\bm{B}$.

\noindent \textbf{Sensitivity analyses.} We performed two sets of sensitive analyses regarding the choice of all the hyperparameters as well as the impact of misspecified tree information. The inference under MMF was relatively robust, in our opinion. Details are provided in Supplementary Material (Section B). In practice, if no prior knowledge is available, we recommend to use the default non-informative prior specification in Sections \ref{sampling}-\ref{prior}.

\section{Real Data} \label{data}

\begin{figure}[!p]
\begin{subfigure}[t]{0.4 \textwidth}
\centering
\includegraphics[width = 1 \textwidth]{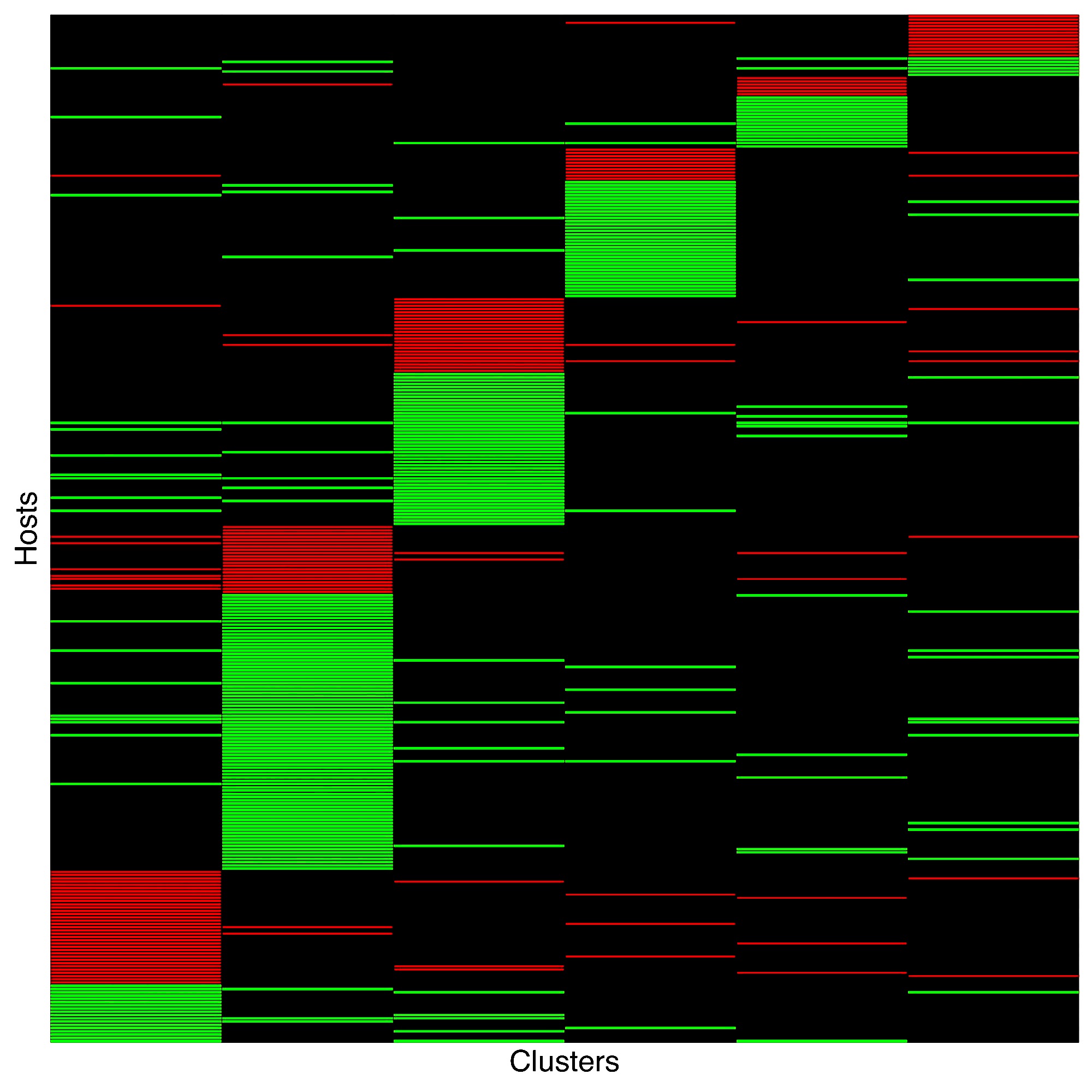}
\caption{Clustering of hosts.}
\label{f4}
\end{subfigure}
\begin{subfigure}[t]{0.4 \textwidth}
\centering
\includegraphics[width = 1 \textwidth]{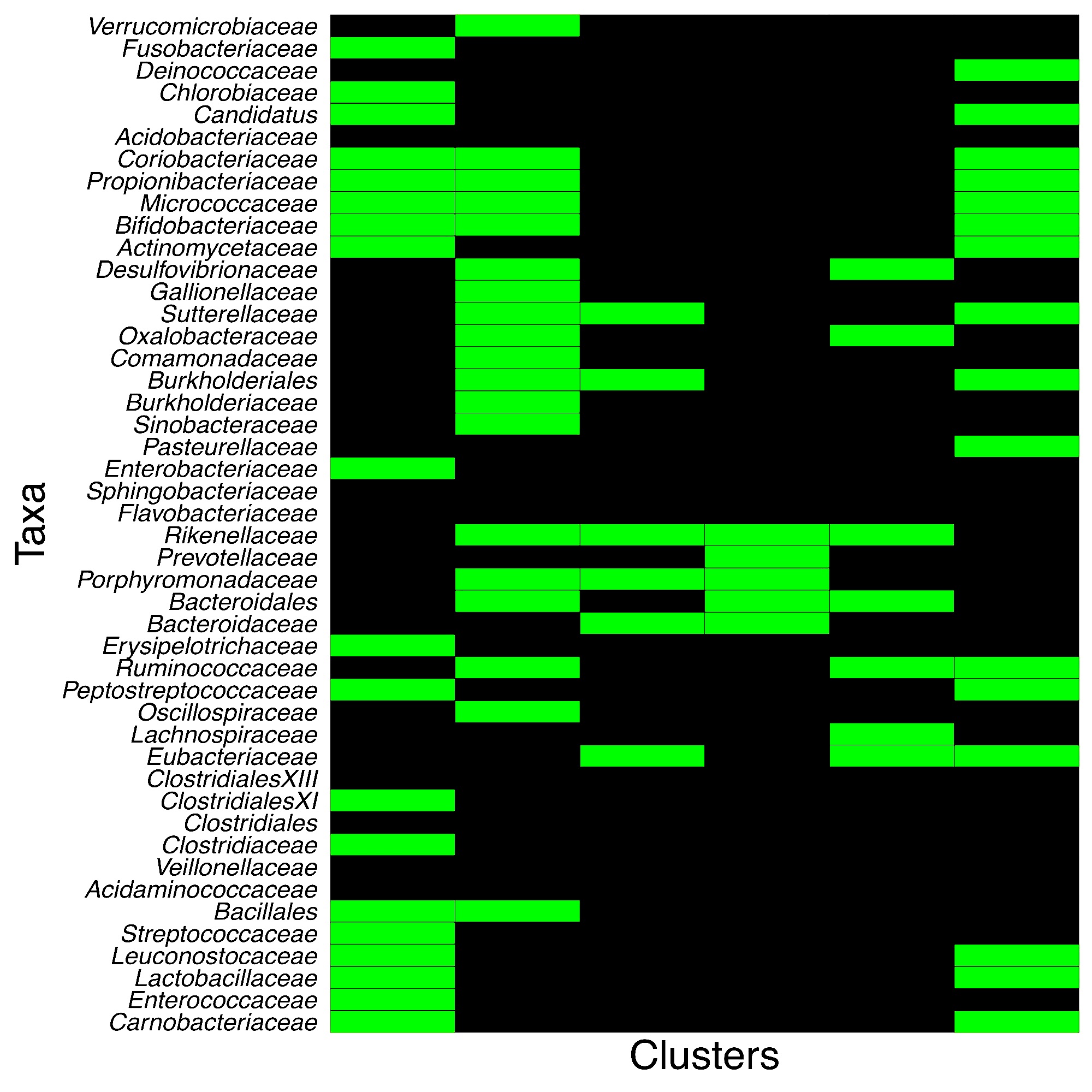}
\caption{Clustering of taxa.}
\end{subfigure}
\begin{subfigure}[t]{0.15 \textwidth}
\centering
\includegraphics[width = 0.885 \textwidth]{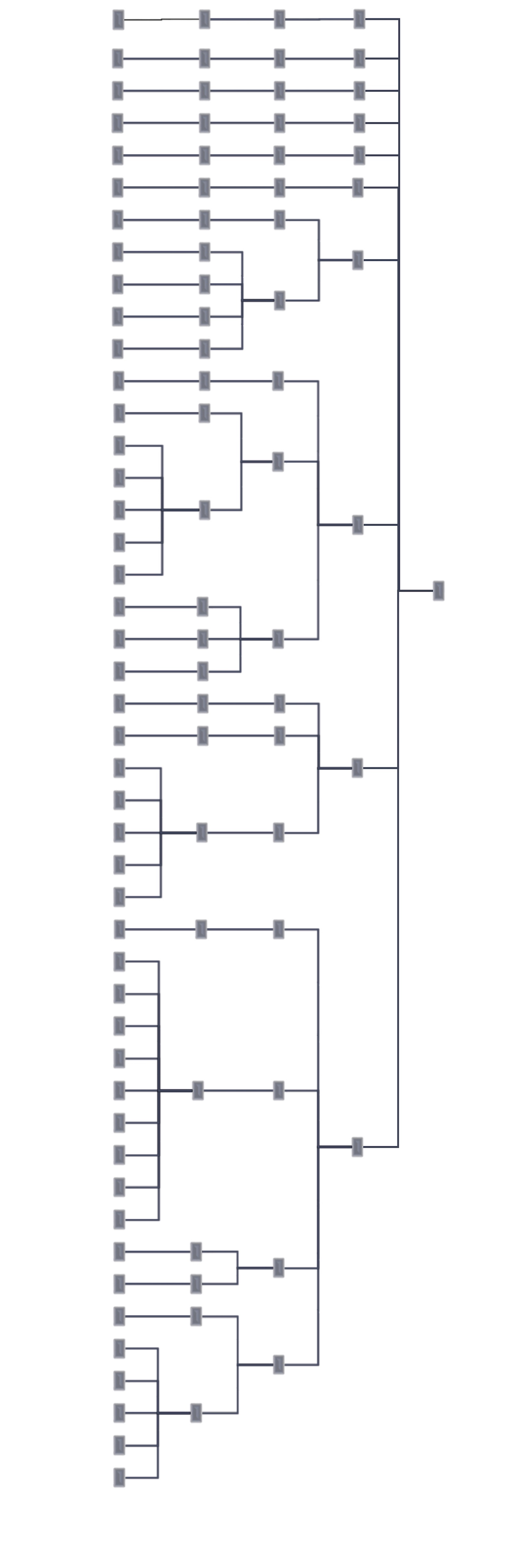}
\end{subfigure}
\caption{Real data. Heatmaps of estimated clusters using the proposed MMF. Colored cells are ones and black cells are zeros. Rows of $\bm{A}$ are hosts arranged in a block-diagonal-liked form. Rows of $\bm{B}$ are taxa which are arranged according to the taxonomic rank tree. Each column represents a cluster with overlaps. The red/green cells in the heatmap of $\bm{A}$ represents patients/controls with ones.}
\label{s3}
\end{figure}

Gut microbes actively interact with their hosts and have profound relevance to inflammatory bowel disease (IBD) which is a very heterogeneous disease at the microbiome level. The goal of this case study was to investigate the heterogeneous microbial profiles in relation to IBD in an unsupervised data-driven manner. We applied the proposed MMF to an IBD microbiome dataset \citep{qin2010human}. The data were obtained by sequencing fecal specimens collected from IBD patients as well as healthy adult controls using the Illumina's Genome Analyzer (metagenomic sequencing); details of the data generating procedure can be found in \cite{qin2010human}. The dataset contained $n = 372$ observations with 240 healthy hosts and 132 IBD patients, and provided information on microbial compositions at various taxonomic levels (kingdom, phylum, class, order, family, genus and species) of these samples. We chose to work with the family level counts because lower levels (e.g., the specie level) had extremely large number of zeros (more than 80\% elements in the count matrix were zeros). In addition, we filtered out families that appear in less than 10\% of samples (i.e. taxa with more than 90\% of zeros) and preserved only families belonging to kingdom bacteria. The resulting data had $p = 46$ taxa for subsequent analysis. The relationships of taxa can be naturally represented by a taxonomic rank tree. Taxa that are closer on the tree tend to have similar activities and functions. We depict the tree of 46 taxa along with their higher taxonomic ranks, kingdom, phylum, class, and order in Figure \ref{s3}. This taxonomic rank tree was used as prior information to encourage the clustering of taxa that are taxonomically similar.

We ran two separate Markov chains of MMF for 10,000 iterations. The first 5,000 iterations were discarded as burn-in and posterior samples were retained every 5th iteration after burn-in. It took 8.2 hours on a 2.3 GHz Quad-Core Intel Core i7 laptop. To monitor the MCMC convergence, we computed the Gelman and Rubin's potential scale reduction factor \citep[PSRF,][]{gelman1992inference} for key parameters. The MCMC diagnostic did not show a sign of lack of convergence: the PSRF was 1.01 for number $K$ of clusters and the median PSRF was $< 1.1$ (with stdev 0.1) for $c_j + \sum_{k = 1}^K a_{ik} w_{jk} b_{jk}$, the quantity on the right-hand side of \eqref{logistic}. The Monte Carlo samples from the two Markov chains were combined for subsequent analysis.

To check the model fit adequacy (measure of ``lack-of-fit"), we performed within-sample prediction that compared the observed composition (i.e. $\bm{x}_i / N_i$) with the posterior predictive mean. The scatter plot of predicted versus observed relative abundance of taxa is given in Figure S.2 (a) in the Supplementary Material showing that the within-sample prediction was accurate. The correlation between two matrices was 0.94, which indicated an adequate model fit.

Figure S.2 (b) in the Supplementary Material shows the posterior distribution of the number $K$ of clusters. The posterior mode occurred at $K = 6$. Conditional on $K$, the posterior estimates of $\bm{A}$ and $\bm{B}$ are shown in Figure \ref{s3}. In Figure \ref{f4}, black cells are 0, green cells are 1 for controls, and red cells are 1 for patients. Samples that did not belong to any clusters are omitted in the figures.

Cluster 1 contained predominantly IBD patients ($\sim$70\%). The biclustering nature of the proposed MMF allowed us to investigate the corresponding subset of taxa that were related to these IBD patients.  For example, the cluster 1 contains family \emph{Enterobacteriaceae}, part of class \emph{Gammaproteobacteria}, which have been reported to increase in relative abundance in patient with IBD \citep{lupp2007host}. The fact that it exclusively belonged to patient-dominated cluster 1 is consistent with its biological relevance to IBD. Moreover, genus \emph{Fusobacterium}, a member of the family \emph{Fusobacteriaceae}, have been found to be at a higher abundance in patients with ulcerative colitis (UC, a subtype of IBD) relative to control subjects \citep{ohkusa2002fusobacterium}. \emph{Fusobacteriaceae} family was also contained in cluster 1 only, which again signified the importance of this family with relevance to IBD. Generally, phyla \emph{Proteobacteria} and \emph{Actinobacteria} are expected to increase in IBD patients \citep{matsuoka2015gut}, which is consistent with our result in cluster 1. Apart from the findings that were confirmed by the existing literature, cluster 1 includes some families in phylum \emph{Firmicutes}, which are known to play major anti-inflammatory roles and therefore their abundances are expected to decrease in IBD patients. Further biological investigation is required to validate this new finding.

Likewise, most hosts in cluster 6 were IBD patients as well. This cluster shared quite a few taxa with cluster 1, which was not surprising as they both contained predominantly IBD patients. However, they also had distinct taxa that are biologically meaningful.
On the one hand, cluster 6 uniquely contained the family \emph{Pasteurellaceae}, of which the abundances tend to increase in patients with Crohn's disease \citep[CD,][]{gevers2014treatment}, another subtype of IBD. This suggested the possibility of these patients belonging to the CD subtype. On the other hand, cluster 1, as discussed earlier, uniquely contained the family \emph{Fusobacteriaceae}, which suggested the possibility of these patients belonging to the UC subtype \citep{ohkusa2002fusobacterium}.

Cluster 2 was dominated by control samples (healthy hosts). It was associated with  families \emph{Bifidobacteriaceae} and \emph{Ruminococcaceae}. Their members, genera \emph{Bifidobacterium} and \emph{Faecalibacterium}, have been shown to be protective of the host from inflammation via several mechanisms \citep{sokol2008faecalibacterium}, including the stimulation of the anti-inflammatory cytokine and down-regulation of inflammatory cytokines. A reduced abundance of genus \emph{Odoribacter}, which belongs to family \emph{Porphyromonadaceae}, has been discovered in the most severe form of UC called pancolitis \citep{morgan2012dysfunction}. It also contained families in class \emph{Betaproteobacteria}, whose relationship with IBD is yet to be established.

Clusters 3, 4, and 5 had a mix of patients and controls. They contained  families \emph{Bacteroidaceae} and \emph{Lachnospiraceae}. Their members, genera \emph{Bacteroides} and \emph{Roseburia}, have been shown to decrease in IBD patients \citep{machiels2014decrease, zhou2016lower}.

We have reported results that were confirmed by the biological literature. Our biclustering results also provided novel insights  into the relationships between microbial abundances and IBD, which need to be further verified by biological experiments. Our discoveries were potentially useful as a guidance to design and conduct more targeted and focused experiments.

For comparison, we applied MMF without the tree information (i.e., using the ordinary IBP prior) to this dataset. The result is shown in Figure S.3.
It identified 4 clusters, and most clusters were dominated by control samples. Without tree information, we failed to identify the cluster associated with IBD patients and two IBD-related bacteria families \emph{Enterobacteriaceae} and \emph{Fusobacteriaceae}, which were successfully discovered when prior knowledge regarding the taxonomic ranks were incorporated in the analysis. 
Additionally, taxa from the same cluster were much less similar taxonomically: the log probability of generating this matrix from the taxonomic tree was -119.95, whereas the log probability of generating the matrix inferred from the pIBP was -91.48, which indicated that the results from the pIBP prior were substantially more consistent with the taxonomic rank tree. Without using the tree information, the lack of taxonomic similarity within the identified clusters made it hard to interpret the results biologically. Although pIBP imposed structures on taxa only, the interpretation of the clusters of hosts was significantly enhanced as we have demonstrated earlier.

\begin{figure}[!p]
\centering
\begin{subfigure}[t]{0.49 \textwidth}
\centering
\includegraphics[width = 1 \textwidth]{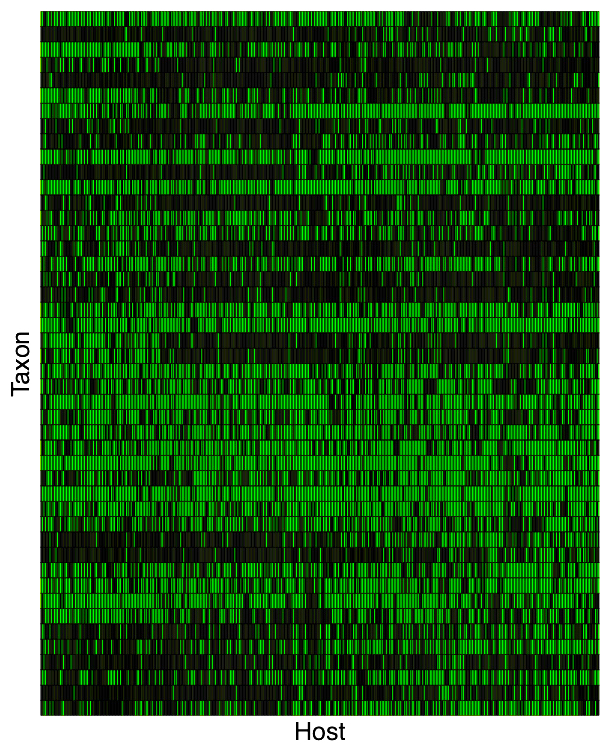}
\caption{MMF}
\end{subfigure}
\begin{subfigure}[t]{0.49 \textwidth}
\centering
\includegraphics[width = 1 \textwidth]{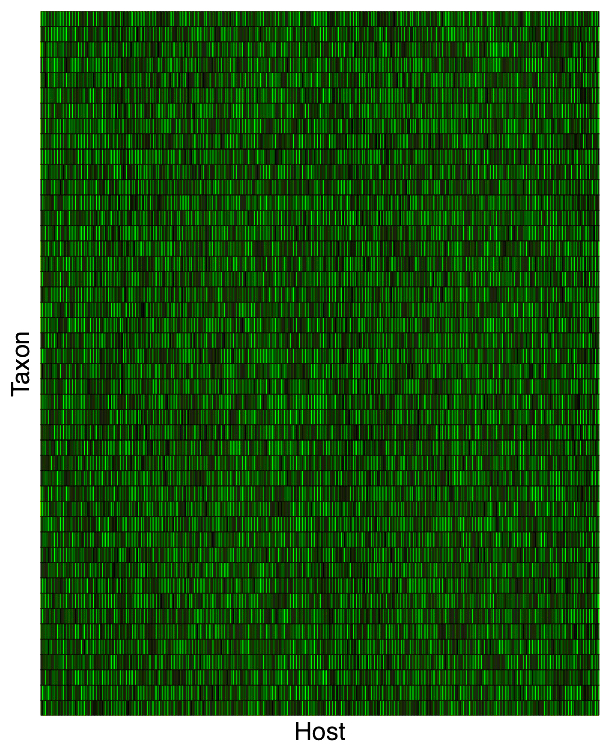}
\caption{MMF with host-specific $s_{ij}$ and $t_{ij}$}
\end{subfigure}
\begin{subfigure}[t]{0.49 \textwidth}
\centering
\includegraphics[width = 1 \textwidth]{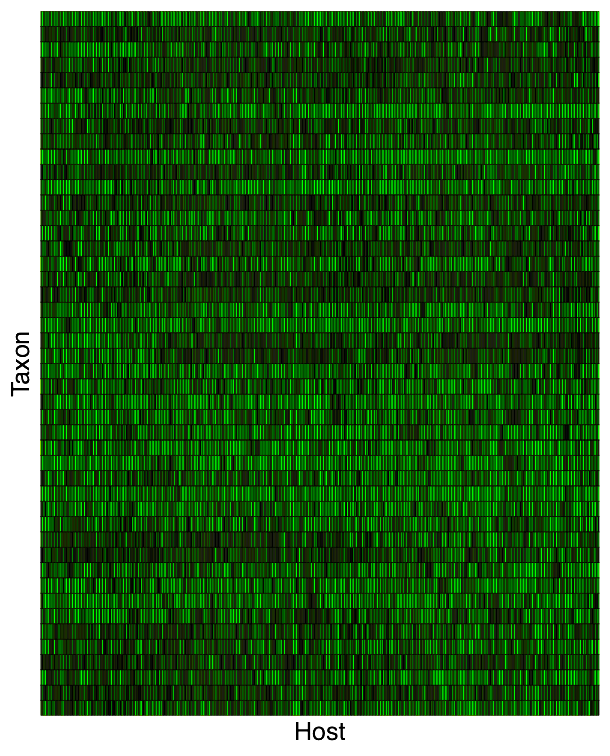}
\caption{MMF with independent Bernoulli prior on $\bm{Z}$}
\end{subfigure}
\begin{subfigure}[t]{0.49 \textwidth}
\centering
\includegraphics[width = 1 \textwidth]{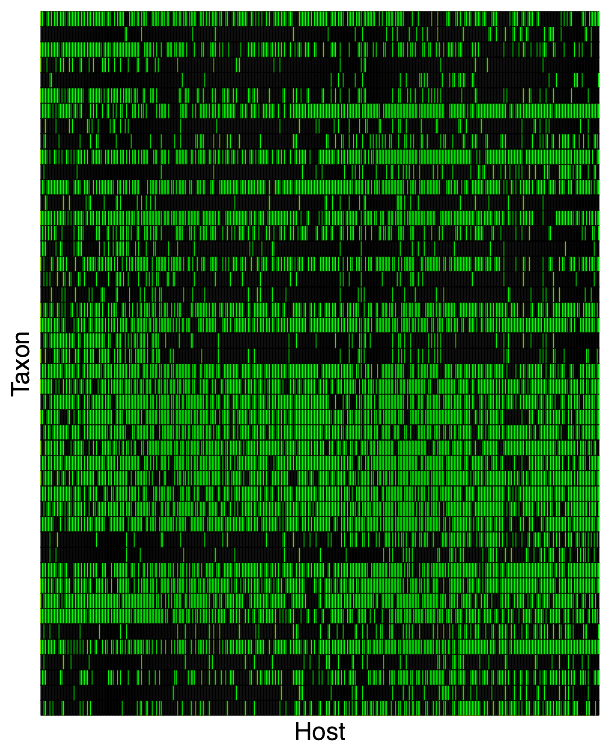}
\caption{Hard thresholded data}
\end{subfigure}
\caption{The posterior mean of the latent binary matrix $\bm{Z}$ obtained from (a)  MMF, (b) modified MMF with host-specific $s_{ij}$ and $t_{ij}$, and (c) modified MMF with independent Bernoulli prior on $\bm{Z}$. The hard thresholded data are given in (d). Colors change from black to green indicating 0 to 1.}
\label{s5}
\end{figure}

As suggested by an anonymous referee, in Figure \ref{s5}, we report the posterior mean of $\bm{Z}$ from (a) the proposed MMF, (b) modified MMF with host-specific $s_{ij}$ and $t_{ij}$ rather than $s_j$ and $t_j $,  and (c) modified MMF with independent Bernoulli prior on $\bm{Z}$ instead of pIBP. Additionally, as a reference, we also plot the thresholded data in Figure \ref{s5} (d) by following the rule in  \cite{cai2019differential}. While there is no gold standard in unsupervised learning, we found that the posterior mean of $\bm{Z}$ from the proposed MMF captures the latent abundance pattern better than (b) and (c) by comparing with the deterministic reference (d).

\section{Discussion} \label{discussion}

In this paper, we have developed a novel identifiable sparse MMF method to simultaneously cluster microbes and hosts. The proposed approach accounts for the compositional, sparse, heterogeneous, and noisy nature of microbiome data, and describes the data generating process by a hierarchical Bayesian model, which allows for probabilistic characterization of latent structures (i.e., overlapping clusters) through full posterior inference. The incorporation of taxonomic knowledge can facilitate the interpretability and reproducibility of the inferred clusters when the prior information resembles the truth. Our simulation results demonstrate the advantage of utilizing prior information to assist inference on latent clusters. In analyzing a human gut microbiome dataset, we find latent microbial communities that are closely related to IBD and its subtypes.  

There are four directions that can be taken to extend this work. First, zero-inflation exists in other data types such as single-cell RNA-seq data. It is far less common to treat single-cell data as multinomial counts and therefore the proposed MMF cannot be directly applied. However, with a minor modification of the sampling distribution (e.g. zero-inflated Poisson distribution), the method can be generalized for biclustering single-cell data. Second, the joint modeling approach can be used for many other tasks beyond matrix factorization. For example, microbial networks can be inferred by replacing the matrix factorization model with a graphical model such as Markov random fields and Bayesian networks on the latent binary indicators $\bm{Z}$. Third, MCMC allows for full posterior inference but is not scalable to large and high-dimensional data. The current inference algorithm can be substantially accelerated by using consensus Monte Carlo algorithms for big-data clustering \citep{ni2019scalable,ni2020consensus} without sacrificing much accuracy. Fourth, the overlapping clusters can be restricted to non-overlapping clusters if desired by considering random partition models including various extensions of the Dirichlet process \citep{lijoi2007controlling, favaro2013mcmc, de2013gibbs}.

\section{Software}

R code, together with a complete documentation, is available on
request from the first author (fangtingzhou@tamu.edu). The data that support the findings of this study are openly available in the R package \texttt{curatedMetagenomicData} which can be downloaded at \url{https://github.com/waldronlab/curatedMetagenomicData/}.

\section*{Supplementary material}

Supplementary material is available online at \url{http://biostatistics.oxfordjournals.org}.

\section*{Acknowledgments}
He's work was partially supported by National Natural Science Foundation of China (No.11801560). Chapkin's research was partially supported by Texas AgriLife Research, the Sid Kyle Chair Endowment, the Allen Endowed Chair in Nutrition \& Chronic Disease Prevention, the Cancer Prevention Research Institute of Texas (RP160589), and the National Institutes of Health (R01-ES025713, R01-CA202697, and R35-CA197707).

\bibliographystyle{biorefs}
\bibliography{ref}

\begin{thebibliography}{99}

\bibitem[Benhadou \emph{and others}(2018)Benhadou, Mintoff, Schnebert and
  Thio]{benhadou2018psoriasis}
\textsc{Benhadou, Farida, Mintoff, Dillon, Schnebert, Benjamin and Thio,
  Hok~Bing}. (2018).
\newblock Psoriasis and microbiota: a systematic review.
\newblock {\em Diseases\/}~\textbf{6}(2), 47.

\bibitem[Bezdek \emph{and others}(1984)Bezdek, Ehrlich and Full]{bezdek1984fcm}
\textsc{Bezdek, James~C, Ehrlich, Robert and Full, William}. (1984).
\newblock {FCM}: the fuzzy c-means clustering algorithm.
\newblock {\em Computers \& Geosciences\/}~\textbf{10}(2-3), 191--203.

\bibitem[Bhattacharya and Dunson(2011)Bhattacharya and
  Dunson]{bhattacharya2011sparse}
\textsc{Bhattacharya, Anirban and Dunson, David~B}. (2011).
\newblock Sparse {B}ayesian infinite factor models.
\newblock {\em Biometrika\/}~\textbf{98}(2), 291--306.

\bibitem[Bishop(2006)Bishop]{bishop2006pattern}
\textsc{Bishop, Christopher}. (2006).
\newblock {\em Pattern recognition and machine learning\/}. Springer.

\bibitem[Buccianti(2013)Buccianti]{buccianti2013compositional}
\textsc{Buccianti, Antonella}. (2013).
\newblock Is compositional data analysis a way to see beyond the illusion?
\newblock {\em Computers \& Geosciences\/}~\textbf{50}, 165--173.

\bibitem[Cai \emph{and others}(2019)Cai, Li, Ma and Xia]{cai2019differential}
\textsc{Cai, Tony, Li, Hongzhe, Ma, Jing and Xia, Yin}. (2019).
\newblock Differential {M}arkov random field analysis with an application to
  detecting differential microbial community networks.
\newblock {\em Biometrika\/}~\textbf{106}(2), 401--416.

\bibitem[Cai \emph{and others}(2017)Cai, Gu and Kenney]{cai2017learning}
\textsc{Cai, Yun, Gu, Hong and Kenney, Toby}. (2017).
\newblock Learning microbial community structures with supervised and
  unsupervised non-negative matrix factorization.
\newblock {\em Microbiome\/}~\textbf{5}, 110.

\bibitem[Cao \emph{and others}(2019{\em a})Cao, Lin and Li]{cao2019large}
\textsc{Cao, Yuanpei, Lin, Wei and Li, Hongzhe}. (2019{\em a}).
\newblock Large covariance estimation for compositional data via
  composition-adjusted thresholding.
\newblock {\em Journal of the American Statistical
  Association\/}~\textbf{114}(526), 759--772.

\bibitem[Cao \emph{and others}(2019{\em b})Cao, Zhang and
  Li]{cao2019multisample}
\textsc{Cao, Yuanpei, Zhang, Anru and Li, Hongzhe}. (2019{\em b}).
\newblock Multisample estimation of bacterial composition matrices in
  metagenomics data.
\newblock {\em Biometrika\/}~\textbf{107}(1), 75--92.

\bibitem[Castaner \emph{and others}(2018)Castaner, Goday,
  et~al.]{castaner2018gut}
\textsc{Castaner, Olga, Goday, Albert,   \emph{and others}}. (2018).
\newblock The gut microbiome profile in obesity: a systematic review.
\newblock {\em International Journal of Endocrinology\/}.

\bibitem[Chen and Li(2016)Chen and Li]{chen2016two}
\textsc{Chen, Eric~Z and Li, Hongzhe}. (2016).
\newblock A two-part mixed-effects model for analyzing longitudinal microbiome
  compositional data.
\newblock {\em Bioinformatics\/}~\textbf{32}(17), 2611--2617.

\bibitem[Chen and Li(2013)Chen and Li]{chen2013variable}
\textsc{Chen, Jun and Li, Hongzhe}. (2013).
\newblock Variable selection for sparse {D}irichlet-multinomial regression with
  an application to microbiome data analysis.
\newblock {\em Annals of Applied Statistics\/}~\textbf{7}(1), 418--442.

\bibitem[Chen \emph{and others}(2016)Chen, Gao and Zhao]{chen2016posterior}
\textsc{Chen, Mengjie, Gao, Chao and Zhao, Hongyu}. (2016).
\newblock Posterior contraction rates of the phylogenetic {I}ndian buffet
  processes.
\newblock {\em Bayesian Analysis\/}~\textbf{11}(2), 477--497.

\bibitem[De~Blasi \emph{and others}(2015)De~Blasi, Favaro, Lijoi, Mena,
  Pr{\"u}nster and Ruggiero]{de2013gibbs}
\textsc{De~Blasi, Pierpaolo, Favaro, Stefano, Lijoi, Antonio, Mena,
  Rams{\'e}s~H, Pr{\"u}nster, Igor and Ruggiero, Matteo}. (2015).
\newblock Are {G}ibbs-type priors the most natural generalization of the
  {D}irichlet process?
\newblock {\em IEEE Transactions on Pattern Analysis and Machine
  Intelligence\/}~\textbf{37}, 212--229.

\bibitem[Favaro and Teh(2013)Favaro and Teh]{favaro2013mcmc}
\textsc{Favaro, Stefano and Teh, Yee~Whye}. (2013).
\newblock {MCMC} for normalized random measure mixture models.
\newblock {\em Statistical Science\/}~\textbf{28}(3), 335--359.

\bibitem[Fettweis \emph{and others}(2019)Fettweis, Serrano
  et~al.]{fettweis2019vaginal}
\textsc{Fettweis, Jennifer~M, Serrano, Myrna~G  \emph{and others}}. (2019).
\newblock The vaginal microbiome and preterm birth.
\newblock {\em Nature Medicine\/}~\textbf{25}, 1012--1021.

\bibitem[Franzosa \emph{and others}(2019)Franzosa, Sirota-Madi
  et~al.]{franzosa2019gut}
\textsc{Franzosa, Eric~A, Sirota-Madi, Alexandra  \emph{and others}}. (2019).
\newblock Gut microbiome structure and metabolic activity in inflammatory bowel
  disease.
\newblock {\em Nature Microbiology\/}~\textbf{4}, 293--305.

\bibitem[Friedman and Alm(2012)Friedman and Alm]{friedman2012inferring}
\textsc{Friedman, Jonathan and Alm, Eric~J}. (2012).
\newblock Inferring correlation networks from genomic survey data.
\newblock {\em PLoS Computational Biology\/}~\textbf{8}(9), e1002687.

\bibitem[Gelman and Rubin(1992)Gelman and Rubin]{gelman1992inference}
\textsc{Gelman, Andrew and Rubin, Donald~B}. (1992).
\newblock Inference from iterative simulation using multiple sequences.
\newblock {\em Statistical Science\/}~\textbf{7}(4), 457--472.

\bibitem[Gevers \emph{and others}(2014)Gevers, Kugathasan
  et~al.]{gevers2014treatment}
\textsc{Gevers, Dirk, Kugathasan, Subra  \emph{and others}}. (2014).
\newblock The treatment-naive microbiome in new-onset {C}rohn's disease.
\newblock {\em Cell Host \& Microbe\/}~\textbf{15}(3), 382--392.

\bibitem[Gloor \emph{and others}(2017)Gloor, Macklaim
  et~al.]{gloor2017microbiome}
\textsc{Gloor, Gregory~B, Macklaim, Jean~M  \emph{and others}}. (2017).
\newblock Microbiome datasets are compositional: and this is not optional.
\newblock {\em Frontiers in Microbiology\/}.

\bibitem[Gopalan \emph{and others}(2014)Gopalan, Ruiz, Ranganath and
  Blei]{gopalan2014bayesian}
\textsc{Gopalan, Prem, Ruiz, Francisco J~R, Ranganath, Rajesh and Blei,
  David~M}. (2014).
\newblock Bayesian nonparametric {P}oisson factorization for recommendation
  systems.
\newblock In:  {\em Proceedings of the Seventeenth International Conference on
  Artificial Intelligence and Statistics\/}, Volume~33. pp.\  275--283.

\bibitem[Grantham \emph{and others}(2020)Grantham, Guan, Reich, Borer and
  Gross]{grantham2020mimix}
\textsc{Grantham, Neal~S, Guan, Yawen, Reich, Brian~J, Borer, Elizabeth~T and
  Gross, Kevin}. (2020).
\newblock Mimix: a bayesian mixed-effects model for microbiome data from
  designed experiments.
\newblock {\em Journal of the American Statistical Association\/}~\textbf{115},
  599--609.

\bibitem[Griffiths and Ghahramani(2005)Griffiths and
  Ghahramani]{griffiths2005infinite}
\textsc{Griffiths, Thomas~L and Ghahramani, Zoubin}. (2005).
\newblock Infinite latent feature models and the {I}ndian buffet process.
\newblock In:  {\em Proceedings of the 18th International Conference on Neural
  Information Processing Systems\/}. pp.\  475--482.

\bibitem[Hoyer(2004)Hoyer]{hoyer2004non}
\textsc{Hoyer, Patrik~O}. (2004).
\newblock Non-negative matrix factorization with sparseness constraints.
\newblock {\em Journal of Machine Learning Research\/}~\textbf{5}, 1457--1469.

\bibitem[Koren \emph{and others}(2013)Koren, Knights, Gonzalez, Waldron,
  Segata, Knight, Huttenhower and Ley]{koren2013guide}
\textsc{Koren, Omry, Knights, Dan, Gonzalez, Antonio, Waldron, Levi, Segata,
  Nicola, Knight, Rob, Huttenhower, Curtis and Ley, Ruth~E}. (2013).
\newblock A guide to enterotypes across the human body: meta-analysis of
  microbial community structures in human microbiome datasets.
\newblock {\em PLoS Computational Biology\/}~\textbf{9}(1), e1002863.

\bibitem[Lahti \emph{and others}(2014)Lahti, Saloj{\"a}rvi, Salonen, Scheffer
  and De~Vos]{lahti2014tipping}
\textsc{Lahti, Leo, Saloj{\"a}rvi, Jarkko, Salonen, Anne, Scheffer, Marten and
  De~Vos, Willem~M}. (2014).
\newblock Tipping elements in the human intestinal ecosystem.
\newblock {\em Nature communications\/}~\textbf{5}, 4344.

\bibitem[Lee and Seung(2000)Lee and Seung]{lee2000algorithms}
\textsc{Lee, Daniel~D and Seung, H~Sebastian}. (2000).
\newblock Algorithms for non-negative matrix factorization.
\newblock In:  {\em Algorithms for non-negative matrix factorization\/}. pp.\
  535--541.

\bibitem[Lijoi \emph{and others}(2007)Lijoi, Mena and
  Pr{\"u}nster]{lijoi2007controlling}
\textsc{Lijoi, Antonio, Mena, Rams{\'e}s~H and Pr{\"u}nster, Igor}. (2007).
\newblock Controlling the reinforcement in {B}ayesian non-parametric mixture
  models.
\newblock {\em Journal of the Royal Statistical Society: Series
  B\/}~\textbf{69}(4), 715--740.

\bibitem[Lin \emph{and others}(2014)Lin, Shi, Feng and Li]{lin2014variable}
\textsc{Lin, Wei, Shi, Pixu, Feng, Rui and Li, Hongzhe}. (2014).
\newblock Variable selection in regression with compositional covariates.
\newblock {\em Biometrika\/}~\textbf{101}(4), 785--797.

\bibitem[Liu \emph{and others}(2019)Liu, Shih, Strawderman, Zhang, Johnson and
  Chai]{liu2019statistical}
\textsc{Liu, Lei, Shih, Ya-Chen~Tina, Strawderman, Robert~L, Zhang, Daowen,
  Johnson, Bankole~A and Chai, Haitao}. (2019).
\newblock Statistical analysis of zero-inflated nonnegative continuous data: a
  review.
\newblock {\em Statistical Science\/}~\textbf{34}(2), 253--279.

\bibitem[Lloyd-Price \emph{and others}(2019)Lloyd-Price, Arze
  et~al.]{lloyd2019multi}
\textsc{Lloyd-Price, Jason, Arze, Cesar  \emph{and others}}. (2019).
\newblock Multi-omics of the gut microbial ecosystem in inflammatory bowel
  diseases.
\newblock {\em Nature\/}~\textbf{569}, 655--662.

\bibitem[Lupp \emph{and others}(2007)Lupp, Robertson  et~al.]{lupp2007host}
\textsc{Lupp, Claudia, Robertson, Marilyn~L  \emph{and others}}. (2007).
\newblock Host-mediated inflammation disrupts the intestinal microbiota and
  promotes the overgrowth of enterobacteriaceae.
\newblock {\em Cell Host \& Microbe\/}~\textbf{2}(2), 119--129.

\bibitem[Machiels \emph{and others}(2014)Machiels, Joossens
  et~al.]{machiels2014decrease}
\textsc{Machiels, Kathleen, Joossens, Marie  \emph{and others}}. (2014).
\newblock A decrease of the butyrate-producing species {R}oseburia hominis and
  {F}aecalibacterium prausnitzii defines dysbiosis in patients with ulcerative
  colitis.
\newblock {\em Gut\/}~\textbf{63}(8), 1275--1283.

\bibitem[Matsuoka and Kanai(2015)Matsuoka and Kanai]{matsuoka2015gut}
\textsc{Matsuoka, Katsuyoshi and Kanai, Takanori}. (2015).
\newblock The gut microbiota and inflammatory bowel disease.
\newblock In:  {\em Seminars in Immunopathology\/}, Volume~37. pp.\  47--55.

\bibitem[Meeds \emph{and others}(2007)Meeds, Ghahramani, Neal and
  Roweis]{meeds2007modeling}
\textsc{Meeds, Edward, Ghahramani, Zoubin, Neal, Radford~M and Roweis, Sam~T}.
  (2007).
\newblock Modeling dyadic data with binary latent factors.
\newblock In:  {\em Proceedings of the 19th International Conference on Neural
  Information Processing Systems\/}. pp.\  977--984.

\bibitem[Miller \emph{and others}(2008)Miller, Griffiths and
  Jordan]{miller2008the}
\textsc{Miller, Kurt~T, Griffiths, Thomas~L and Jordan, Michael~I}. (2008).
\newblock The phylogenetic {I}ndian buffet process: a non-exchangeable
  nonparametric prior for latent features.
\newblock In:  {\em Proceedings of the 24th Conference on Uncertainty in
  Artificial Intelligence\/}. pp.\  403--410.

\bibitem[Morgan \emph{and others}(2012)Morgan, Tickle
  et~al.]{morgan2012dysfunction}
\textsc{Morgan, Xochitl~C, Tickle, Timothy~L  \emph{and others}}. (2012).
\newblock Dysfunction of the intestinal microbiome in inflammatory bowel
  disease and treatment.
\newblock {\em Genome Biology\/}~\textbf{13}, R79.

\bibitem[Ni \emph{and others}(2020)Ni, Ji and M{\"u}ller]{ni2020consensus}
\textsc{Ni, Yang, Ji, Yuan and M{\"u}ller, Peter}. (2020).
\newblock Consensus monte carlo for random subsets using shared anchors.
\newblock {\em Journal of Computational and Graphical Statistics\/}, 1--12.

\bibitem[Ni \emph{and others}(2019{\em a})Ni, M{\"u}ller, Diesendruck,
  Williamson, Zhu and Ji]{ni2019scalable}
\textsc{Ni, Yang, M{\"u}ller, Peter, Diesendruck, Maurice, Williamson, Sinead,
  Zhu, Yitan and Ji, Yuan}. (2019{\em a}).
\newblock Scalable {B}ayesian nonparametric clustering and classification.
\newblock {\em Journal of Computational and Graphical
  Statistics\/}~\textbf{29}, 53--65.

\bibitem[Ni \emph{and others}(2019{\em b})Ni, M{\"u}ller and
  Ji]{ni2019bayesian}
\textsc{Ni, Yang, M{\"u}ller, Peter and Ji, Yuan}. (2019{\em b}).
\newblock Bayesian double feature allocation for phenotyping with electronic
  health records.
\newblock {\em Journal of the American Statistical Association\/}, 1--15.

\bibitem[Ohkusa \emph{and others}(2002)Ohkusa, Sato
  et~al.]{ohkusa2002fusobacterium}
\textsc{Ohkusa, Toshifumi, Sato, Nobuhiro  \emph{and others}}. (2002).
\newblock Fusobacterium varium localized in the colonic mucosa of patients with
  ulcerative colitis stimulates species-specific antibody.
\newblock {\em Journal of Gastroenterology and Hepatology\/}~\textbf{17}(8),
  849--853.

\bibitem[Parmigiani \emph{and others}(2002)Parmigiani, Garrett, Anbazhagan and
  Gabrielson]{parmigiani2002statistical}
\textsc{Parmigiani, Giovanni, Garrett, Elizabeth~S, Anbazhagan, Ramaswamy and
  Gabrielson, Edward}. (2002).
\newblock A statistical framework for expression-based molecular classification
  in cancer.
\newblock {\em Journal of the Royal Statistical Society: Series
  B\/}~\textbf{64}, 717--736.

\bibitem[Qin \emph{and others}(2010)Qin, Li  et~al.]{qin2010human}
\textsc{Qin, Junjie, Li, Ruiqiang  \emph{and others}}. (2010).
\newblock A human gut microbial gene catalogue established by metagenomic
  sequencing.
\newblock {\em Nature\/}~\textbf{464}, 59--65.

\bibitem[Ren \emph{and others}(2017)Ren, Bacallado, Favaro, Holmes and
  Trippa]{ren2017bayesian}
\textsc{Ren, Boyu, Bacallado, Sergio, Favaro, Stefano, Holmes, Susan and
  Trippa, Lorenzo}. (2017).
\newblock Bayesian nonparametric ordination for the analysis of microbial
  communities.
\newblock {\em Journal of the American Statistical Association\/}~\textbf{112},
  1430--1442.

\bibitem[Ro{\v{c}}kov{\'a} and George(2016)Ro{\v{c}}kov{\'a} and
  George]{rovckova2016fast}
\textsc{Ro{\v{c}}kov{\'a}, Veronika and George, Edward~I}. (2016).
\newblock Fast {B}ayesian factor analysis via automatic rotations to sparsity.
\newblock {\em Journal of the American Statistical
  Association\/}~\textbf{111}(516), 1608--1622.

\bibitem[Shafiei \emph{and others}(2015)Shafiei, Dunn, Boon, MacDonald, Walsh,
  Gu and Bielawski]{shafiei2015biomico}
\textsc{Shafiei, Mahdi, Dunn, Katherine~A, Boon, Eva, MacDonald, Shelley~M,
  Walsh, David~A, Gu, Hong and Bielawski, Joseph~P}. (2015).
\newblock Biomico: a supervised bayesian model for inference of microbial
  community structure.
\newblock {\em Microbiome\/}~\textbf{3}, 8.

\bibitem[Shi \emph{and others}(2016)Shi, Zhang and Li]{shi2016regression}
\textsc{Shi, Pixu, Zhang, Anru and Li, Hongzhe}. (2016).
\newblock Regression analysis for microbiome compositional data.
\newblock {\em The Annals of Applied Statistics\/}~\textbf{10}(2), 1019--1040.

\bibitem[Sokol \emph{and others}(2008)Sokol  et~al.]{sokol2008faecalibacterium}
\textsc{Sokol, Harry  \emph{and others}}. (2008).
\newblock Faecalibacterium prausnitzii is an anti-inflammatory commensal
  bacterium identified by gut microbiota analysis of {C}rohn disease patients.
\newblock {\em Proceedings of the National Academy of
  Sciences\/}~\textbf{105}(43), 16731--16736.

\bibitem[Tilg and Moschen(2014)Tilg and Moschen]{tilg2014microbiota}
\textsc{Tilg, Herbert and Moschen, Alexander~R}. (2014).
\newblock Microbiota and diabetes: an evolving relationship.
\newblock {\em Gut\/}~\textbf{63}(9), 1513--1521.

\bibitem[Turnbaugh \emph{and others}(2007)Turnbaugh, Ley
  et~al.]{turnbaugh2007human}
\textsc{Turnbaugh, Peter~J, Ley, Ruth~E  \emph{and others}}. (2007).
\newblock The human microbiome project.
\newblock {\em Nature\/}~\textbf{449}, 804--810.

\bibitem[Wadsworth \emph{and others}(2017)Wadsworth, Argiento, Guindani,
  Galloway-Pena, Shelburne and Vannucci]{wadsworth2017integrative}
\textsc{Wadsworth, W~Duncan, Argiento, Raffaele, Guindani, Michele,
  Galloway-Pena, Jessica, Shelburne, Samuel~A and Vannucci, Marina}. (2017).
\newblock An integrative bayesian dirichlet-multinomial regression model for
  the analysis of taxonomic abundances in microbiome data.
\newblock {\em BMC Bioinformatics\/}~\textbf{18}, 94.

\bibitem[Watts \emph{and others}(2018)Watts, Ritchie, Inouye and
  Holt]{watts2018fastspar}
\textsc{Watts, Stephen~C, Ritchie, Scott~C, Inouye, Michael and Holt,
  Kathryn~E}. (2018).
\newblock {FastSpar}: rapid and scalable correlation estimation for
  compositional data.
\newblock {\em Bioinformatics\/}~\textbf{35}(6), 1064--1066.

\bibitem[Wood \emph{and others}(2006)Wood, Griffiths and Ghahramani]{wood2006a}
\textsc{Wood, Frank, Griffiths, Thomas~L and Ghahramani, Zoubin}. (2006).
\newblock A non-parametric bayesian method for inferring hidden causes.
\newblock In:  {\em Proceedings of the Twenty-Second Conference on Uncertainty
  in Artificial Intelligence\/}. pp.\  536--543.

\bibitem[Wu \emph{and others}(2019)Wu, Casciola-Rosen, Rosen and
  Zeger]{wu2019barlcm}
\textsc{Wu, Zhenke, Casciola-Rosen, Livia, Rosen, Antony and Zeger, Scott~L}.
  (2019).
\newblock A {B}ayesian approach to restricted latent class models for
  scientifically-structured clustering of multivariate binary outcomes.
\newblock {\em bioRxiv\/}.

\bibitem[Xia \emph{and others}(2013)Xia, Chen, Fung and Li]{xia2013logistic}
\textsc{Xia, Fan, Chen, Jun, Fung, Wing~Kam and Li, Hongzhe}. (2013).
\newblock A logistic normal multinomial regression model for microbiome
  compositional data analysis.
\newblock {\em Biometrics\/}~\textbf{69}, 1053--1063.

\bibitem[Xu \emph{and others}(2020)Xu, Demmer and Li]{xu2020zero}
\textsc{Xu, Tianchen, Demmer, Ryan~T. and Li, Gen}. (2020).
\newblock Zero-inflated poisson factor model with application to microbiome
  read counts.
\newblock {\em Biometrics\/}, accepted.

\bibitem[Zhou \emph{and others}(2012)Zhou, Hannah, Dunson and
  Carin]{zhou2012beta}
\textsc{Zhou, Mingyuan, Hannah, Lauren, Dunson, David and Carin, Lawrence}.
  (2012).
\newblock Beta-negative binomial process and {P}oisson factor analysis.
\newblock In:  {\em Proceedings of the Fifteenth International Conference on
  Artificial Intelligence and Statistics\/}. pp.\  1462--1471.

\bibitem[Zhou and Zhi(2016)Zhou and Zhi]{zhou2016lower}
\textsc{Zhou, Yingting and Zhi, Fachao}. (2016).
\newblock Lower level of bacteroides in the gut microbiota is associated with
  inflammatory bowel disease: a meta-analysis.
\newblock {\em BioMed Research International\/}.

\end{thebibliography}

% \begin{table}[!p]
% \centering
% \caption{Summary of the complete hierarchical model. }
% \begin{tabular}{ll}
% Multinomial sampling &$\bm{x}_i \sim \mbox{Multinomial}(N_i, \bm{\pi}_i) \mbox{~~with~~} \bm{\pi}_i = {\bm{\gamma}_i} / {\sum_{j = 1}^p\gamma_{ij}}$ \\\hline
% Mixture representation &$\gamma_{ij} \sim I(z_{ij} = 1) \mbox{Gamma}(s_j, 1) + I(z_{ij} = 0) \mbox{Gamma}(t_j, 1)$ \\
% of the composition &$(s_j, t_j) \sim p(s_j, t_j) = \mathrm{Gamma}(s_j | \alpha_s, \beta_s)$ \\
% & $ \qquad \qquad\qquad\qquad\qquad\times \mathrm{Gamma}(t_j | \alpha_t, \beta_t) \times I(s_j > t_j)$ \\\hline
% Latent matrix &$\mathrm{logit}\{\Pr(z_{ij} = 1)\} = c_j + \sum_{k = 1}^K a_{ik} w_{jk} b_{jk}$\\
% factorization &$\bm{B} \sim \mbox{pIBP}(m), \quad a_{ik} \sim \mbox{Bernoulli}(\rho), \quad \rho \sim \mbox{Beta}(\alpha_\rho, \beta_\rho)$\\
% &$w_{jk} \sim \mathrm{Gamma} (\alpha_w, \beta_w), \quad c_j \sim \mathrm{N}(\mu_c, \sigma_c)$
% \end{tabular}
% \label{shm}
% \end{table}

\begin{table}[!p] 
\centering
\caption{Simulation results of the proposed MMF and competing methods. Average errors in estimating $\bm{A}$ and $\bm{B}$ are quantified as the Hamming distance between the estimated and true matrices, normalized by the respective total number of elements. The numbers in the parentheses are standard deviations. The smallest errors are in boldface. The competing methods are low rank approximation (LRA), non-negative matrix factorization (NNMF), zero-inflated Poisson factor model (ZIPFM), and two-step multinomial matrix factorization (TSMF). }
\begin{tabular}{ccccccc}
\toprule
\multirow{2}{*}{$(s, t)$}  & \multicolumn{2}{c}{(2, 0.7)} & \multicolumn{2}{c}{(3, 0.6)} & \multicolumn{2}{c}{(5, 0.5)} \cr
\cmidrule(lr){2-3} \cmidrule(lr){4-5} \cmidrule(lr){6-7} 
& Error $\bm{A} $& Error $\bm{B}$ & Error $\bm{A} $& Error $\bm{B}$ & Error $\bm{A} $& Error $\bm{B}$ \cr
\midrule
MMF & \tabincell{c}{0.373 \\ (0.062)} & \tabincell{c}{\textbf{0.167} \\ (0.029)}  & \tabincell{c}{\textbf{0.171} \\ (0.022)} & \tabincell{c}{\textbf{0.055} \\ (0.021)} & \tabincell{c}{\textbf{0.117} \\ (0.023)} & \tabincell{c}{\textbf{0.057} \\ (0.028)}\\
\midrule
LRA & \tabincell{c}{\textbf{0.298} \\ (0.042)} & \tabincell{c}{0.269 \\ (0.023)} & \tabincell{c}{0.205 \\ (0.052)} & \tabincell{c}{0.185 \\ (0.017)} & \tabincell{c}{0.203 \\ (0.058)} & \tabincell{c}{0.165 \\ (0.021)}  \\
\midrule
NNMF & \tabincell{c}{0.351 \\ (0.049)} & \tabincell{c}{0.279 \\ (0.030)} & \tabincell{c}{0.288 \\ (0.062)} & \tabincell{c}{0.247 \\ (0.021)} & \tabincell{c}{0.256 \\ (0.059)} & \tabincell{c}{0.208 \\ (0.014)}  \\
\midrule
ZIPFM &  \tabincell{c}{0.425 \\ (0.014)} & \tabincell{c}{0.258 \\ (0.031)} & \tabincell{c}{0.291 \\ (0.008)} & \tabincell{c}{0.249 \\ (0.023)} & \tabincell{c}{0.246 \\ (0.002)} & \tabincell{c}{0.232 \\ (0.022)} \\
\midrule
TSMF & \tabincell{c}{0.382 \\ (0.092)} & \tabincell{c}{0.253 \\ (0.033)} & \tabincell{c}{0.325 \\ (0.057)} & \tabincell{c}{0.132 \\ (0.043)} & \tabincell{c}{0.237 \\ (0.092)} & \tabincell{c}{0.089 \\ (0.018)} \\
\bottomrule
\end{tabular} 
\label{t1}
\end{table}

\begin{table}[!p] 
\centering
\caption{Simulation results of the misspecified model. Average errors in estimating $\bm{A}$ and $\bm{B}$ are quantified as the Hamming distance between the estimated and true matrices, normalized by the respective total number of elements. The numbers in the parentheses are standard deviations. The competing methods are low rank approximation (LRA), non-negative matrix factorization (NNMF), zero-inflated Poisson factor model (ZIPFM), and two-step multinomial matrix factorization (TSMF). }
\begin{tabular}{cccccc}
\toprule
& MMF & LRA & NNMF & ZIPFM & TSMF \cr
\midrule
Error rate $\bm{A}$ & \tabincell{c}{0.323 \\ (0.072)} & \tabincell{c}{0.301 \\ (0.044)} & \tabincell{c}{0.351 \\ (0.062)} & \tabincell{c}{0.368 \\ (0.015)} & \tabincell{c}{0.329 \\ (0.083)} \cr
\midrule
Error rate $\bm{B}$ & \tabincell{c}{0.157 \\ (0.036)} & \tabincell{c}{0.293 \\ (0.032)} & \tabincell{c}{0.319 \\ (0.023)} & \tabincell{c}{0.382 \\ (0.019)} & \tabincell{c}{0.191 \\ (0.047)} \cr
\bottomrule
\end{tabular}
\label{t2}
\end{table}

\end{document}